# Wave propagation in single column woodpile phononic crystals: Formation of tunable band gaps

**Eunho Kim**

William E. Boeing Department of Aeronautics & Astronautics, University of Washington

211 Guggenheim Hall, Seattle, WA 98195-2400, USA

eunhokim80@gmail.com

**Jinkyu Yang** [1)]

William E. Boeing Department of Aeronautics & Astronautics, University of Washington

311B Guggenheim Hall, Seattle, WA 98195-2400, USA

jkyang@aa.washington.edu

## Abstract

We study the formation of frequency band gaps in single column woodpile phononic crystals composed of orthogonally stacked slender cylinders. We focus on investigating the effect of the cylinders' local vibrations on the dispersion of elastic waves along the stacking direction of the woodpile phononic crystals. We experimentally verify that their frequency band structures depend significantly on the bending resonant behavior of unit cells. We propose a simple theoretical model based on a discrete element method to associate the behavior of locally resonant cylindrical rods with the band gap formation mechanism in woodpile phononic crystals. The findings in this work imply that we can achieve versatile control of frequency band structures in phononic crystals by using woodpile architectures. The woodpile phononic crystals can form a new type of vibration filtering devices that offer an enhanced degree of freedom in manipulating stress wave propagation.

[1] Corresponding Author.
Tel.: 2065436612 ; Fax: 2065430217
E-mail: jkyang@aa.washington.edu

# 1 Introduction

Phononic crystals (PCs) are defined as artificially fabricated structures in periodic architectures. They have received significant attention in the last decade as an emerging technology for controlling elastic and acoustic wave propagation (Kushwaha, 1996; Liu et al., 2000; Vasseur et al., 2001; Hussein, 2007). In particular, PCs can exhibit forbidden frequency bands – called band gaps – with tailored frequency limits. Previous studies have shown the versatility and tunability of frequency band structures in PCs by: (i) manipulating the geometry and mechanical properties of unit-cell elements; (ii) imposing various boundary conditions; and (iii) changing assembling architectures of constituents (Herbold et al., 2009; Boechler et al., 2011; Yang et al., 2012).

Among various types of previously designed PCs, woodpile PCs are assemblies of periodically stacked longitudinal components – often embedded in matrices – in 3D architectures (Jiang et al., 2009; Wu and Chen, 2011). These woodpile PCs can offer an enhanced degree of freedom in adjusting their frequency band structures by modifying design parameters, such as rod spacing, alignment angles, and stacking sequences. By leveraging such controllability, their optical counterpart called woodpile "photonic" crystals has been studied extensively to demonstrate rich behavior of frequency band structures in the electromagnetic spectrum (Lin et al., 1998; Noda et al., 2000). In acoustics, researchers have investigated the woodpile architectures to create band gaps primarily through the interactions with the surrounding fluidic media. For example, Wu and Chen (2011) investigated the formation of acoustic band gaps in woodpile PCs with the focus on modeling pressure field patterns of air around the simple cubic lattices. Jiang et al. (2009) studied the elastic wave propagation in composite structures consisting of hard polymer matrix and rigid woodpile structure for underwater acoustic wave absorption. Most research on woodpile PCs, however, has been based on rigid architectures or simple translational oscillations of unit cell elements without fully considering their development of complex deformation modes.

In this research, we investigate tunable phononic band gaps in single column woodpile PCs made of orthogonally stacked cylinders. The slender rods in woodpile architectures develop bending resonances in low frequency regimes, which can significantly influence the dispersive behavior of elastic waves along the stacking direction of PCs. In principle, such characteristics of woodpile PCs are equivalent to the physical mechanism of locally resonant PCs (i.e., sonic crystals or acoustic meta-materials in a broader sense), that attracted attention due to their capability of forming low frequency band gaps without requiring large lattice constants (Liu et al., 2000; Hirsekorn, 2004; Jiang et al., 2009). Unlike these conventional PCs, however, the woodpile PCs use intrinsic bending modes of unit-cell elements to achieve local resonances, without necessitating complicated architectures of mass-in-mass configurations or multi-layered structures composed of hard and soft materials. Functionally, these simple woodpile PCs can reduce wave dissipation substantially, resulting in distinctive cutoff frequencies and high transmission gains in pass bands.

To investigate the effect of the cylinders' dynamics on the formation of band gaps, we test a series of woodpile PC specimens with different rod lengths and boundary conditions. Based on the natural modes of these cylinders, a simple numerical method using a discrete element model is developed. Despite its simple monoatomic configuration, we observe that woodpile PCs create multiple band gaps whose cutoff frequencies are characterized by the resonant modes of the cylinders. We find that these band gaps can be adjusted by changing the rod lengths or by manipulating pre-compression applied to the woodpile PCs due to the nonlinear Hertzian contact among the cylindrical elements. The formation of surface modes is also detected due to the localized vibrations of cylinders at the boundaries. The mode shapes of such propagating and localized waves are investigated by using the discrete element model, and the theoretical band structures of the woodpile PCs are found to be in good agreement with the experimental measurements.

The rest of the paper is structured as follows: First, we describe the experimental setup in Sec. 2. We then introduce a theoretical model of the woodpile PCs in Sec. 3. Based on this model, we develop a numerical method in Sec. 4 to calculate frequency band structures. Sec. 5 describes a comparison between analytical, numerical, and experimental results. Lastly, in Sec. 6, we conclude the paper with summary and future work.

## 2 Experimental approach

The overall configuration of the experimental setup is shown in Fig. 1. The assembled woodpile PCs consist of vertically stacked 23 cylindrical rods, whose centers form orthogonal contacts with neighboring elements. The rods are made of fused quartz (density $\rho = 2200\ \text{kg}/m^3$, elastic modulus $E = 72$ GPa, and Poisson's ratio $v = 0.17$). We test five specimens of woodpile PCs with different rod lengths ($L$ = 40, 50, 60, 70, 80 mm), while keeping their diameters identical ($D$ = 5.0 mm). We excite the woodpile PC using a piezoelectric actuator (Piezomechanik PSt 500\10\25) that is in direct contact with the specimen's top element. The actuator is powered by a ±5 V chirp signal (1 Hz ~ 50 kHz), which is generated by a function generator (Agilent 33220A) during a 0.5 second period. We compress the column of woodpile PC using two compression springs (spring coefficient $k$ = 1.49 N/mm) attached on the top of the actuator. To adjust the pre-compression in an accurate manner, we employ a linear stage that allows fine control of the springs' compressive displacements. In this study, we apply the pre-compression in the range between 10.1 N and 39.9 N, where the lowest pre-compression corresponds to the weight of the actuator part. To prevent buckling under compressive loads, the cylindrical members are guided by supporting rods in the test frame. When an excessive amount of pre-compression is applied, we witness that the 1D column becomes noticeably distorted, making the unit cell elements not parallel anymore. Within the range of the pre-compression considered in this study, however, we observe that the cylinders remain well aligned. To measure the transmitted waves through the woodpile PC, we mount a piezoelectric force sensor (PCB 200-B01) at the bottom of the test frame. The measured signals are acquired from an oscilloscope (Agilent MOS7054B) with a sampling frequency of 100 kHz. We repeat each testing five times to ensure repeatability of testing results.

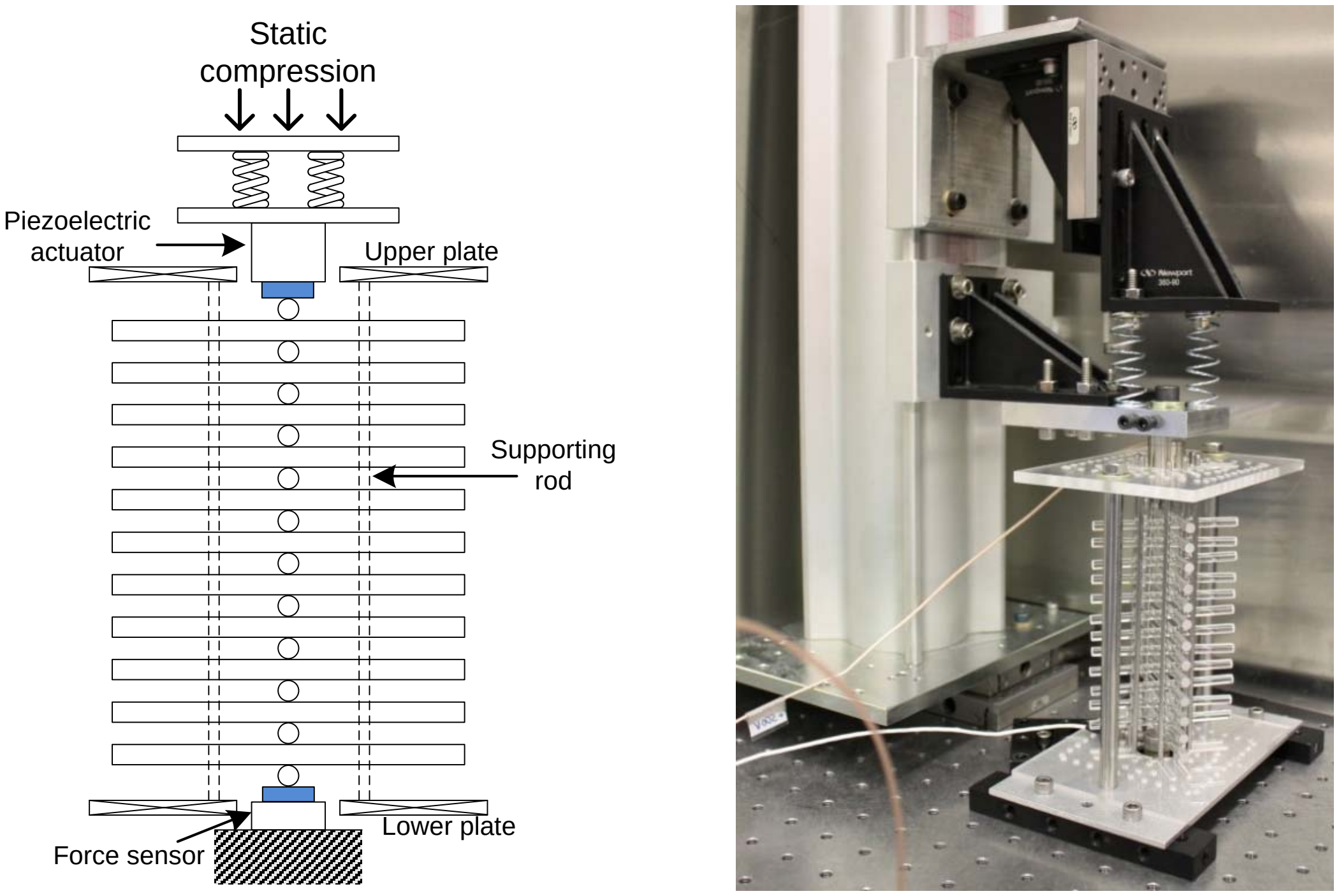

(a) (b)

**Figure 1** (a) Schematic diagram of the test setup. (b) Digital image of the experimental setup.

# 3 Analytical model

## 3.1 Unit cell model

We begin with modeling bending vibrations of a unit cell (i.e., a single cylinder). While a slender rod has an infinite number of resonant modes including torsional and extension-compression modes, its primary vibration motions are governed dominantly by bending vibration modes in a low frequency domain. In this study, we obtain natural frequencies and corresponding mode shapes of slender rods using a finite element method (FEM). The natural frequencies of bending vibration modes within 50 kHz are listed in Table 1 for various lengths of rods considered in this study. To verify the FEM results, we conduct modal testing using a piezoelectric sensor bonded on the surface of a cylinder. The test results are in satisfactory agreement with the FEM simulations, showing only 0.74 % error in average (Table 1). The slight discrepancies are probably due to the errors in the material properties and dimensions of the specimen, as well as the bonded sensor on the surface of the cylinder that affects the natural frequency of the cylinders.

**Table 1** Natural frequencies of cylindrical rods under bending vibration modes. The shaded region corresponds to the frequencies beyond the cutoff frequency (50 kHz) to be considered in this study.

| Rod length (mm) | Mode1 (kHz) | | Mode2 (kHz) | | Mode3 (kHz) | | Mode4 (kHz) | | Mode5 (kHz) | |
|---|---|---|---|---|---|---|---|---|---|---|
| | Exp. | FEM | Exp. | FEM | Exp. | FEM | Exp. | FEM | Exp. | FEM |
| 40 | 15.41 | 15.19 | 39.52 | 39.44 | | | | | | |
| 50 | 9.92 | 9.86 | 26.4 | 26.08 | 48.98 | 48.52 | | | | |
| 60 | 6.87 | 6.89 | 18.46 | 18.45 | 35.10 | 34.80 | | | | |
| 70 | 5.19 | 5.09 | 13.73 | 13.72 | 26.40 | 26.11 | 41.81 | 41.62 | | |
| 80 | 3.97 | 3.91 | 10.53 | 10.59 | 20.45 | 20.27 | 32.50 | 32.55 | 47.15 | 47.02 |

According to the FEM analysis, the first four bending modes of a 70 mm-long rod are shown in Fig. 2. The odd-numbered modes are symmetric, while the even-numbered bending modes are anti-symmetric. When a group of cylinders are stacked together with respect to their center-of-mass positions, the anti-symmetric vibration modes of these cylinders do not affect the axial wave propagation along their stacking direction. This is because the nodal point at the contact location does not move during the local vibrations of the cylinders. Thus, we account for only symmetric bending modes in modeling a unit cell.

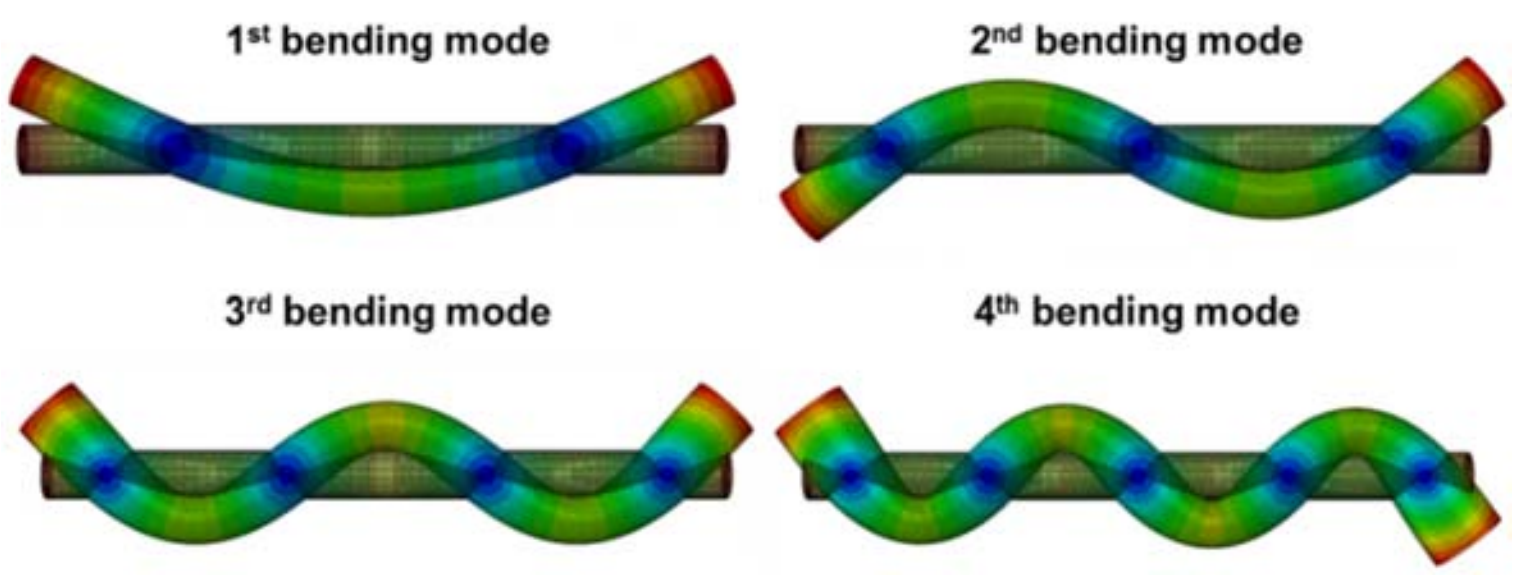

**Figure 2** The first four bending modes of a 70 mm-long rod.

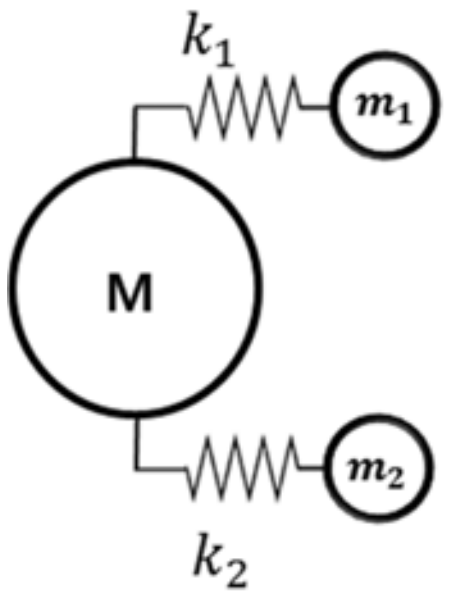

**Figure 3** Discrete element model of a unit cell.

To mimic the cylinder's vibration modes up to the second order (i.e., 1st and 3rd bending modes), we construct a simplified spring-mass model composed of two auxiliary masses ($m_1$ and $m_2$) connected to the primary mass (*M*) through two linear spring ($k_1$ and $k_2$) (Fig. 3). Equations of motion of this spring-mass system are expressed as follows:

$$M\ddot{u} = k_1(v-u) + k_2(w-u),$$
$$m_1\ddot{v} = k_1(u-v),$$
$$m_2\ddot{w} = k_2(u-w). \tag{1}$$

Here, $u$, $v$, and $w$ represent the displacements of masses $M$, $m_1$, and $m_2$, respectively. Under the harmonic oscillations of these masses, we can obtain a characteristic equation of the system as follows:

$$\omega^6 - \left(\frac{1}{M}(k_1+k_2) + \frac{k_1}{m_1} + \frac{k_2}{m_2}\right)\omega^4 + \frac{k_1k_2}{m_1m_2}\left(1 + \frac{m_1+m_2}{M}\right)\omega^2 = 0, \tag{2}$$

where $\omega$ is the angular frequency of the oscillation. We solve Eq. (2) to obtain two resonant frequencies as follows:

$$\omega_{1,2} = \left[\frac{1}{2}\left\{\left(\frac{k_1}{m_1} + \frac{k_2}{m_2} + \frac{1}{M}(k_1+k_2)\right) \mp \sqrt{\left(\frac{k_1}{m_1} + \frac{k_2}{m_2} + \frac{1}{M}(k_1+k_2)\right)^2 - 4\frac{k_1}{m_1}\frac{k_2}{m_2}\left(1 + \frac{m_1+m_2}{M}\right)}\right\}\right]^{1/2} \tag{3}$$

For a short cylinder with a single mode of bending vibration, we can disregard the second spring mass terms $(k_2, m_2)$ in Eq. (2) and (3). In this case, the resonant frequency of the cylinder is reduced to the following form:

$$\omega_1 = \sqrt{\frac{(M+m_1)k_1}{Mm_1}} \tag{4}$$

As shown in Table 1, the 40 mm-long rod encompasses only the first symmetric mode within a 50 kHz range. For the rest of rods tested in this study (up to $L$ = 80 mm), double auxiliary masses are sufficient to simulate the second-order bending vibration.

### 3.2 Periodic woodpile model

To construct a discrete element model for the woodpile PC, we connect the unit cell elements in a row as shown in Fig 4. The mechanical stiffness between two adjacent unit cells can be characterized by the nonlinear Hertzian law of two cylinders under a 90° contact angle. Similarly, we account for the chain's boundary conditions, where the first and last cylinders form a mechanical contact with the tips of the sensor and the actuator, respectively. In all these contacts, we use this linear approximation of contact stiffness between the cylinders, considering that the magnitude of dynamic disturbances are orders-of-magnitude smaller than the static pre-compression (Herbold et al., 2009; Boechler et al., 2011).

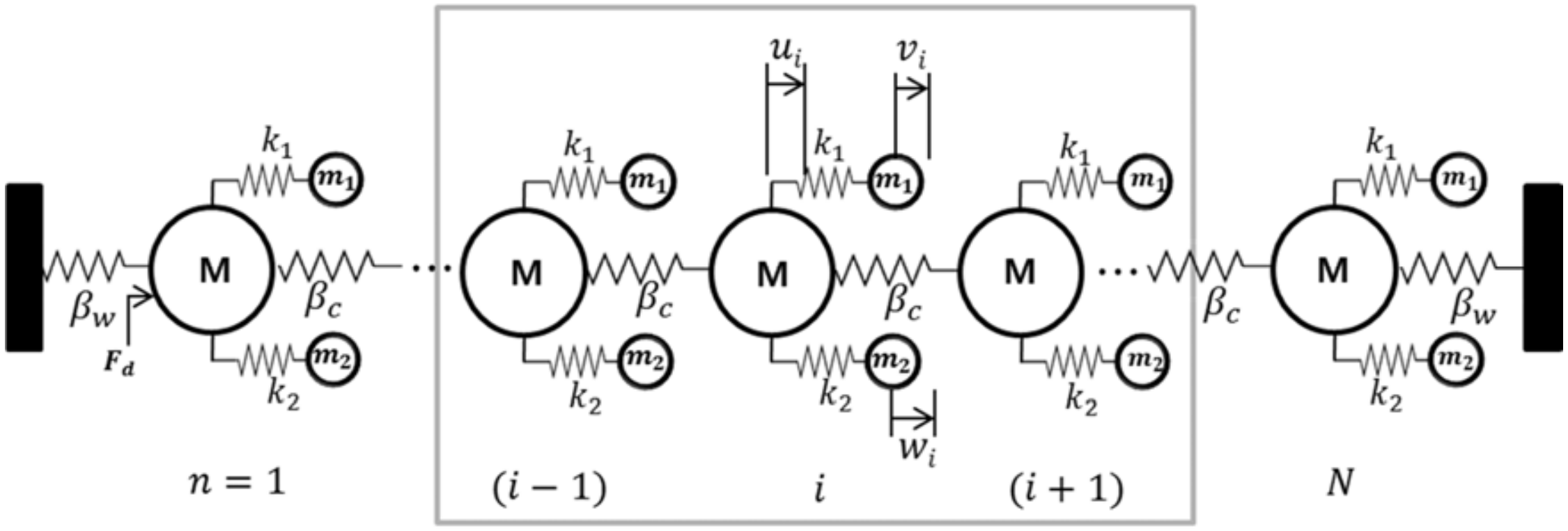

**Figure 4** Discrete element model for a chain of slender cylindrical rods.

Let the linearized contact stiffness between the rods be $\beta_c$, while the contact stiffness between the wall and the rod be $\beta_w$. After the linearization of the Hertzian contact theory under static pre-compression $F_0$, the stiffness values can be expressed as follows (Herbold et al., 2009):

$$\beta_c = \frac{3}{2} A_c{}^{2/3} F_0^{1/3},$$

$$\beta_w = \frac{3}{2} A_w{}^{2/3} F_0^{1/3}. \quad (5)$$

Here $A_c$ is the Hertzian contact coefficient in the chain defined as $A_c \equiv \frac{2E_c\sqrt{R_c}}{3(1-v_c{}^2)}$, where $E_c$, $v_c$, and $R_c$ are Young's modulus, Poisson's ratio, and radius of the rod, respectively (Johnson, 1985). $A_w$ denotes the contact coefficient at the wall boundary, which is defined as $A_w \equiv \frac{2\sqrt{D_w B_1}}{3(V_w+V_c)} \cdot \frac{1}{B_2^{1.5}}$ for the contact between a cylindrical element and a curved wall (Puttock and Thwaite, 1969). Here, $D_w$ is a diameter of the wall curvature (which is 0.60 m for the round-shaped sensor cap in this study), and $V_w$ and $V_c$ are expressed as $V_w \equiv \frac{1-v_w^2}{\pi E_w}$ and $V_c \equiv \frac{1-v_c^2}{\pi E_c}$, where $E_w$ and $v_w$ are Young's modulus and Poisson's ratio of the boundary ($E_w$ = 200 GPa and $v_w$ = 0.29 for the sensor cap made of steel). $B_1$ and $B_2$ are defined as $\frac{-1}{e}\frac{dE(e)}{de}$ and $K(e)$, where $K(e)$ and $E(e)$ are the complete elliptic integral of the first and second kind respectively, and $e$ is the eccentricity of the elliptic contact area ($e$ = 0 for the 90° contact angle in this study). Given these geometrical and material parameters of the system, we obtain $B_1$ = 3.40 and $B_2$ = 4.45.

In this study, we limit the contact angle to 90°, but it should be noted that the change of contact angles between neighboring cylinders allows for the variation of contact stiffness. For example, we can achieve a stiffer chain of cylinders by reducing their contact angles from 90 ° to a smaller angle. In dynamic environments, this can provide tunability of propagating waves' speed and frequency band structures as reported in the references (Khatri et al., 2012; Li et al., 2012).

Based on the linearized stiffness values, the equations of motion of the *i*-th unit cell (see the boxed area in Fig. 4) are:

$$M_i\ddot{u}_i = \beta_c(u_{i-1} - 2u_i + u_{i+1}) + k_1(v_i - u_i) + k_2(w_i - u_i), \ \ i \in \{2, \ldots, N-1\},$$
$$m_1\ddot{v}_i = k_1(u_i - v_i), \ \ i \in \{1, \ldots, N\}$$
$$m_2\ddot{w}_i = k_2(u_i - w_i), i \in \{1, \ldots, N\}. \quad (6)$$

To account for the boundary conditions, the first equation in Eq. (6) is modified for the first and last unit cells as follows:

$$M_1\ddot{u}_1 = F_\mathrm{d} - \beta_w u_1 - \beta_c(u_1 - u_2) + k_1(v_1 - u_1) + k_2(w_1 - u_1),$$
$$M_N\ddot{u}_N = -\beta_w u_N + \beta_c(u_{N-1} - u_N) + k_1(v_N - u_N) + k_2(w_N - u_N). \quad (7)$$

Here $F_{\mathrm{d}}$ is the dynamic force applied to the first cylinder by the piezoelectric actuator.

We first derive an analytical solution of these wave equations under the assumption of an infinite chain of cylinders, while we will attempt to integrate them numerically in the next Section. We impose the Floquet theory (Brillouin, 1953), postulating that the harmonic motions of particles in a certain cell are identical to those in its neighboring cells except for a phase change corresponding to a spatial distance $a$ between adjacent particles:

$$u_i = Ue^{i(\omega t-kx)} \rightarrow u_{i\pm1} = Ue^{i(\omega t-k(x\pm a))} = u_i e^{\mp ika}. \tag{8}$$

Here $k$ is the wave number, ω is the angular frequency, $U$ is the wave amplitude of the primary mass $M$, and $a = 2R - \delta_0$ is the equilibrium unit cell length under the initial compression $\delta_0$ caused by the static pre-compression. We plug Eq. (8) into Eq. (6) to obtain the following dispersion relationship:

$$\omega^6 - \left\{\left(\frac{k_1}{m_1}+\frac{k_2}{m_2}\right)+\frac{2\beta_c}{M}(1-\cos(ka))+\frac{1}{M}(k_1+k_2)\right\}\omega^4 + \left\{\frac{k_1k_2}{m_1m_2}+\frac{2\beta_c}{M}(1-\cos(ka))\left(\frac{k_1}{m_1}+\frac{k_2}{m_2}\right)+\frac{k_1k_2}{M}\left(\frac{1}{m_1}+\frac{1}{m_2}\right)\right\}\omega^2 - \frac{2\beta_c}{M}\frac{k_1k_2}{m_1m_2}(1-\cos(ka)) = 0 \tag{9}$$

We obtain dispersion curves (solid curves in Fig. 5 (a)) in the case of 70-mm rods by solving Eq. (9). We observe that the band structure of the woodpile PC develops multiple, distinctive pass bands. This is in sharp contrast to the creation of a single pass-band structure in a monoatomic chain composed of short cylindrical particles (Li et al., 2012). In Fig. 5 (a), it should be noted that the onset frequencies of the second and third pass-bands are identical to the resonant frequencies of the unit cell model as expressed in Eq. (3) (i.e., $f_{i2} = {}^{\omega_1}/_{2\pi}$, $f_{i3} = {}^{\omega_2}/_{2\pi}$). This is plausible mathematically because when the wave number is zero ($k$ = 0), the dispersion relation in Eq. (9) becomes exactly the same as the characteristic equation of the unit cell (Eq. (2)). From a physical viewpoint, this means that when the wave length approaches infinite, the neighboring cells exhibit identical motions, preventing the force interactions between the adjacent cells. This leads to the oscillations of each cylinder with its own resonant frequencies. Therefore, these local resonances initiate the formation of additional pass bands, resulting in multiple band gaps. Considering that local resonant modes are simulated by auxiliary masses, we also find that the number of additional pass bands in woodpile PCs is identical to the added degree of freedom in DEM (i.e., the number of auxiliary masses).

For comparison purposes, Fig. 5(a) includes the dispersion curve (dashed line) of the same woodpile PC structure under the assumption of no local resonances (i.e., infinite values of $k_1$ and $k_2$). As mentioned previously, this non-resonating model introduces only a single dispersion curve without developing any additional wave modes in the frequency range of our interest. Notably, the cutoff frequency of this non-resonant structure (marked as $f_{c,n}$ in Fig. 5 (a)) is positioned higher than the cutoff frequency of the first band gap ($f_{c1}$) developed in the locally resonant structure (mathematically, $f_{c,n} = \sqrt{4\beta_c/(M + m_1 + m_2)} > f_{c1}$).

To further investigate the effect of local resonances on the positions of frequency bandgaps, we intentionally change the resonant frequencies of a 70-mm cylinder and calculate the shift of frequency bandgaps numerically (Fig. 5 (b)). More specifically, the two resonant frequencies, $f_{i2}$ and $f_{i3}$, are varied simultaneously by a factor of $p$ by altering the spring coefficients, $k_1$and $k_2$. The corresponding local resonant frequencies are expressed in the abscissa of Fig. 5 (b) after being normalized by the nominal resonant frequencies given in Table 1. Here, small $p$ values represent low resonant frequencies, while large $p$ values denote high resonant frequencies ($p$ = 1 is the condition of the nominal case shown in Fig. 5 (a)). We observe that as the resonant frequencies increase, the band gaps (grey zones in Fig. 5 (b)) become wider and shift to higher frequencies. It is noteworthy that the upper boundaries of these bandgaps (dotted lines in Fig. 5 (b)) are straight lines. This confirms that the onset of additional pass bands are attributed to the local resonant frequencies of unit cell elements. Another interesting point is that the lower cutoff frequency of the first pass band ($f_{c1}$) approaches asymptotically to $f_{c,n}$ (black dash line). This explains that locally resonant PCs always generate lower bandgaps than their non-resonating counterparts. This is an interesting feature, which is similar to the principle of acoustic metamaterials that allow the formation of low frequency band gaps below the cutoff frequencies defined by the Bragg's lattice constants (Liu et al., 2000).

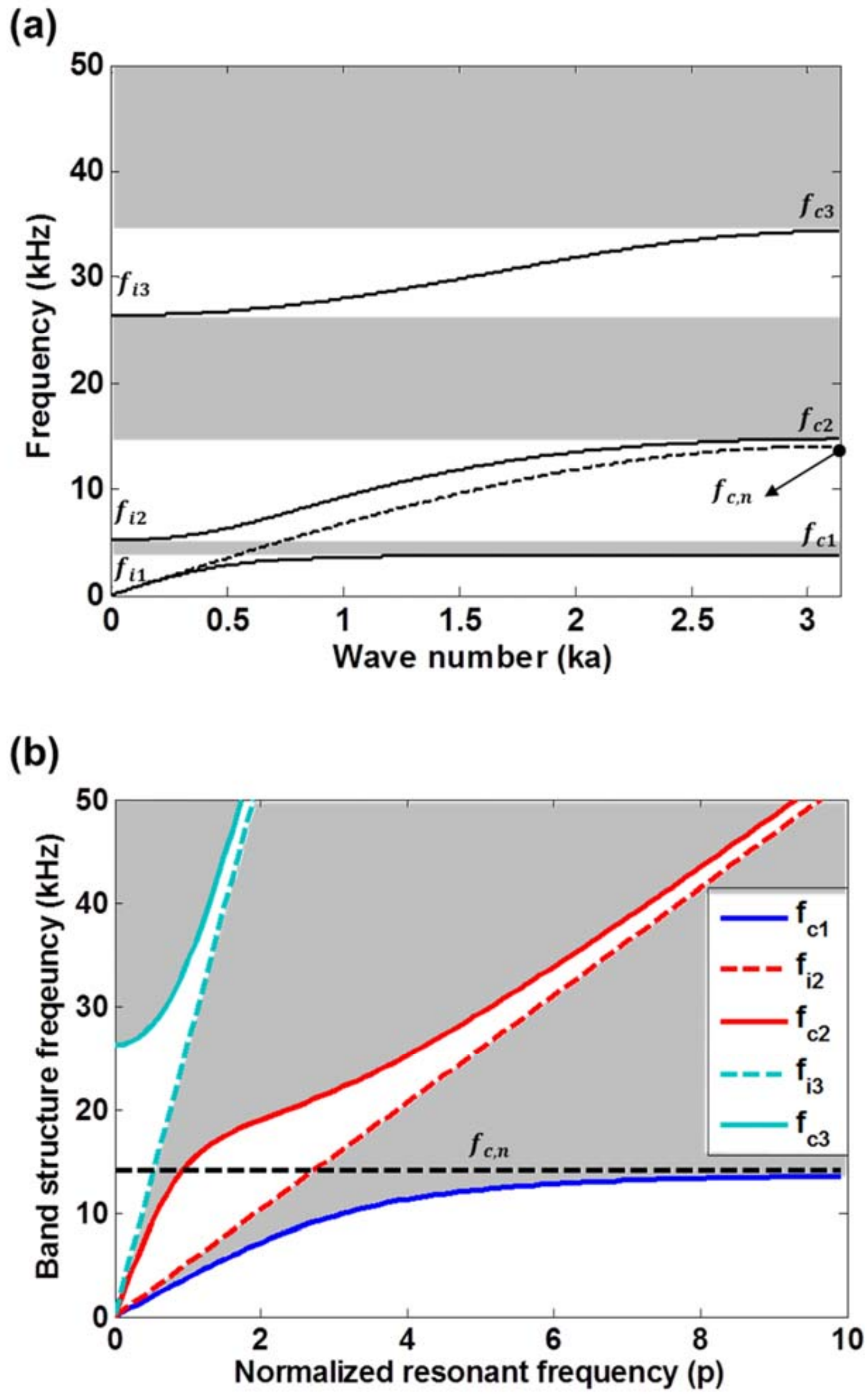

**Figure 5** (a) Dispersion curves of a single column woodpile PC composed of 70 mm-long cylindrical rods. Solid (dashed) curves represent the frequency band structures of a locally resonating (non-resonating) PC. The shaded areas denote forbidden frequency bands, including band gaps. (b) Variations of frequency band structures as a function of normalized resonant frequencies ($p$) of a unit cell. The dashed horizontal line represents the cutoff frequency ($f_{c,n}$) of the band structure in a non-resonating woodpile PC.

## 4 Numerical approach

### 4.1 Parameter determination

Now we need to determine the parameters in the DEM, such as masses (M, $m_1, m_2$) and stiffnesses ($k_1, k_2$). In this study we use an optimization scheme as outlined in Fig. 6. First, we start with assuming a mass distribution (M, $m_1, m_2$), whose total mass is identical to the mass of the cylindrical rod. Given the mass distribution, we calculate the stiffness of springs ($k_1, k_2$) by using a unit cell model in Eq. (3). Under the condition of ${}^{k_1}/_{m_1} < {}^{k_2}/_{m_2}$, the stiffness values $k_1$ and $k_2$ are expressed mathematically as:

$$k_1 = \frac{C_1C_2(\omega_1^2+\omega_2^2)-\sqrt{\left(C_1C_2(\omega_1^2+\omega_2^2)\right)^2-4C_1C_2C_3C_4\omega_1^2\omega_2^2}}{2C_1C_4}$$

$$k_2 = \frac{C_1C_2(\omega_1^2+\omega_2^2)+\sqrt{\left(C_1C_2(\omega_1^2+\omega_2^2)\right)^2-4C_1C_2C_3C_4\omega_1^2\omega_2^2}}{2C_1C_3} \tag{10}$$

where the constants $C_1$, $C_2$, $C_3$, and $C_4$ are defined as follows:

$$C_1 = Mm_1m_{2,}$$

$$C_2 = M+m_1 + m_2,$$

$$C_3 = m_1(M + m_2\,),$$

$$C_4 = m_2(M + m_1\,). \tag{11}$$

Here the cylinder's resonant frequencies $\omega_1$ and $\omega_2$ are given by the FEM simulations (Table 1). The next step is that we calculate the cutoff frequencies of the dispersion curves. In particular, we calculate $\omega_{c2}$ and $\omega_{c3}$ from Eq. (9), which correspond to the upper cutoff frequencies of the two high pass bands (see $f_{c2}$ ($= \omega_{c2}/2\pi$) and $f_{c3}$ ($= \omega_{c3}/2\pi$) in Fig. 5 (a)). We calculate the square error ($E$) of these numerical cutoff frequencies relative to those measured in experiments by:

$$E = \left(\omega_{c2} - \omega_{c2}|_{exp}\right)^2 + \left(\omega_{c3} - \omega_{c3}|_{exp}\right)^2. \tag{12}$$

We repeat this process in various mass distributions until we minimize $E$. We ultimately obtain the optimized parameters of mass and stiffness values that yield the best agreement between numerical and empirical cutoff frequencies. In this calculation, the cutoff frequency of the first pass band ($f_{c1}$) is excluded, because it is not clearly distinguished in the experiment data.

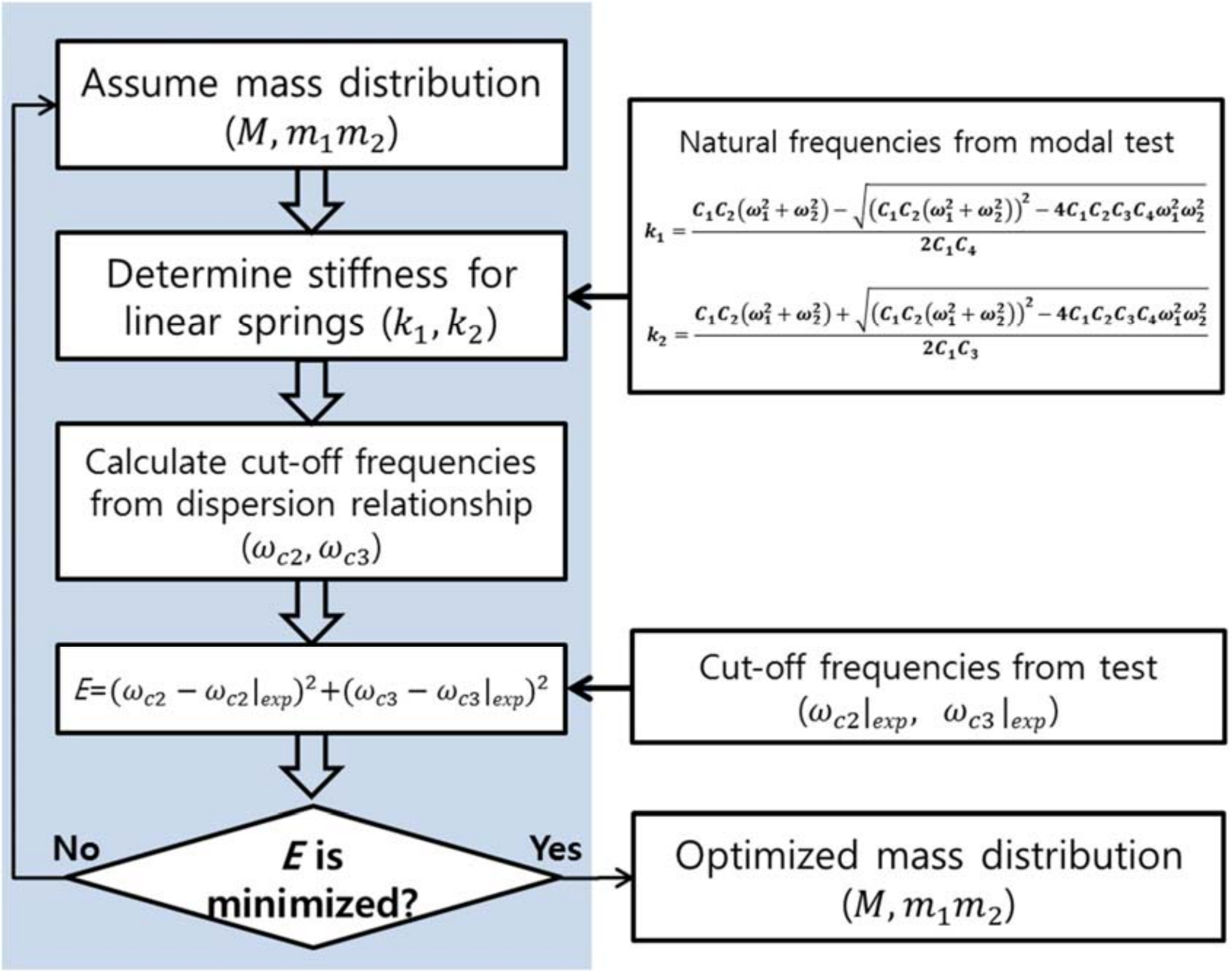

**Figure 6** Flow chart of the optimization process to determine system parameters.

Figure 7 shows the square error map for the 70-mm rod model as a function of the primary ($M$) and auxiliary ($m_1$) masses. We observe that the error function is strongly governed by both parameters. As shown in the magnified figure in the inset, we find that there is a global minimum point at $M$ = 0.850 g and $m_1$ = 1.413 g. We will discuss the optimization results of various lengths of cylindrical rods in Section 5.4.

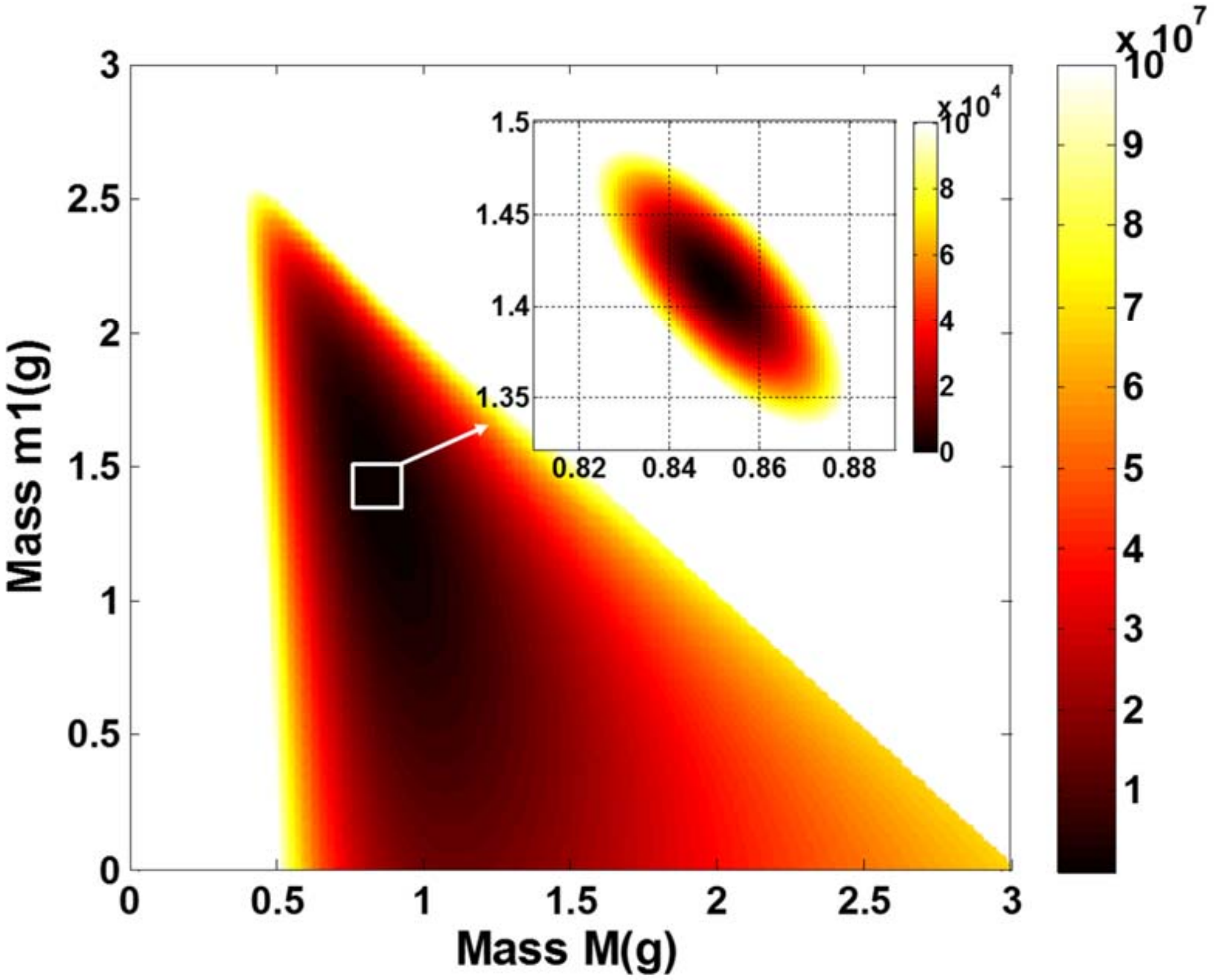

**Figure 7** Square error function ($E$) in terms of primary ($M$) and auxiliary ($m_1$) masses for the 70mm rod model. The color maps represent the magnitude of $E$. The inset shows a magnified view of the error function map around the minimum value.

## 4.2 Frequency responses of a finite chain

Based on the optimized parameters, we investigate the frequency responses of the woodpile PC using the DEM. While we employed Floquet theory for the analysis of an infinite chain, here we use a state-space approach (Franklin et al., 1993; Boechler et al., 2011) to account for the woodpile PC with a finite chain length and various boundary conditions. Compared to the analytical model, this state-space model allows us to calculate the transmission gains of the woodpile PC as a function of input frequencies. Furthermore, various vibrational modes – including surface modes – can be simulated by this finite chain model. The state-space approach based on DEM also saves computational efforts significantly compared to a full 3D FEM, which requires highly refined mesh and time steps to account for nonlinear contact among cylinders.

The governing equations of motion (Eqs. (6) and (7)) can be expressed in a general matrix form of state-space representation as follows (Franklin et al., 1993):

$$\dot{\boldsymbol{X}} = \boldsymbol{A}\boldsymbol{X} + \boldsymbol{B}F_d,$$

$$F_N = \boldsymbol{C}\boldsymbol{X} + \boldsymbol{D}F_d. \tag{13}$$

Here $\boldsymbol{X}$ is the state vector containing displacement and velocity components of each particle, and the matrices $\boldsymbol{A}$, $\boldsymbol{B}$, and $\boldsymbol{C}$ are state, input, and output matrices, respectively. The direct transformation matrix $\boldsymbol{D}$ is a scalar with a zero value because there is no direct input affecting the response of our system. The input of the system is the dynamic disturbance applied to the first particle in the chain ($F_d$ in Fig. 4). The output of the system is $F_N = \beta_w u_N$, which denotes the transmitted contact force at the end of the chain. Mathematically, $\boldsymbol{X}$, $\boldsymbol{A}$, $\boldsymbol{B}$, and $\boldsymbol{C}$ can be expressed as:

$$\boldsymbol{X} = \begin{pmatrix} u_1 \\ \vdots \\ u_N \\ v_1 \\ \vdots \\ v_N \\ w_1 \\ \vdots \\ w_N \\ \dot{u}_1 \\ \vdots \\ \vdots \\ \dot{w}_N \end{pmatrix}, \mathbf{A} = \begin{pmatrix} \mathbf{0} & \boldsymbol{I} \\ \boldsymbol{M}^{-1}\boldsymbol{K} & \mathbf{0} \end{pmatrix}, \boldsymbol{B} = \begin{pmatrix} 0 \\ \vdots \\ 0 \\ 0 \\ \vdots \\ 0 \\ 0 \\ \vdots \\ 0 \\ 1/M \\ \vdots \\ \vdots \\ 0 \end{pmatrix}, \ \boldsymbol{C} = \begin{pmatrix} 0 \\ \vdots \\ \beta_w \\ 0 \\ \vdots \\ 0 \\ 0 \\ \vdots \\ 0 \\ 0 \\ \vdots \\ \vdots \\ 0 \end{pmatrix}^T, \tag{14}$$

where $\boldsymbol{M}$and $\boldsymbol{K}$ are mass and stiffness matrices that are determined from the equations of motion in Eqs. (6) and (7), and $\boldsymbol{I}$ is an identity matrix. Accordingly, mass and stiffness matrices have the following forms:

$$\boldsymbol{M} = \begin{pmatrix} \boldsymbol{M}_{11} & \mathbf{0} & \mathbf{0} \\ \mathbf{0} & \boldsymbol{M}_{22} & \mathbf{0} \\ \mathbf{0} & \mathbf{0} & \boldsymbol{M}_{33} \end{pmatrix},$$

$$\boldsymbol{K} = \begin{pmatrix} \boldsymbol{K}_{11} & \boldsymbol{K}_{12} & \boldsymbol{K}_{13} \\ \boldsymbol{K}_{12} & \boldsymbol{K}_{22} & \boldsymbol{K}_{23} \\ \boldsymbol{K}_{13} & \boldsymbol{K}_{23} & \boldsymbol{K}_{33} \end{pmatrix}. \tag{15}$$

Here $\boldsymbol{M_{11}}$, $\boldsymbol{M_{22}}$, and $\boldsymbol{M_{33}}$ are scalar matrices whose diagonal elements are $M$, $m_1$, and $m_2$, respectively. $\boldsymbol{K_{11}}$ is defined as follows:

$$\boldsymbol{K_{11}} = \begin{pmatrix} -\beta_c-\beta_{\mathrm{w}}-k_1-k_2 & \beta_{\mathrm{c}} & & & & \\ \beta_{\mathrm{c}} & -2\beta_{\mathrm{c}}-k_1-k_2 & \beta_{\mathrm{c}} & & \mathbf{0} & \\ & \beta_{\mathrm{c}} & \ddots & & & \\ & & \ddots & \beta_{\mathrm{c}} & & \\ & \mathbf{0} & \beta_{\mathrm{c}} & -2\beta_{\mathrm{c}}-k_1-k_2 & \beta_{\mathrm{c}} \\ & & & \beta_{\mathrm{c}} & -\beta_{\mathrm{c}}-\beta_{\mathrm{w}}-k_1-k_2 \end{pmatrix}, \quad (16)$$

Similar to the mass matrices, $\boldsymbol{K_{12}}$, $\boldsymbol{K_{13}}$, $\boldsymbol{K_{22}}$, $\boldsymbol{K_{23}}$, and $\boldsymbol{K_{33}}$ are scalar matrices with their diagonal values of $k_1$, $k_2$, $-k_1$, 0 and $-k_2$, respectively.

To calculate the transmission gain of dispersive waves in a finite particle chain, we solve Eq. (13) by using MATLAB's bode function to obtain the transfer function of the system (i.e., $F_N/F_d$). Also, we obtain the natural frequencies and mode shapes of the finite chain by solving the eigen-value problem as below:

$$\boldsymbol{M}\ddot{\boldsymbol{X}} + \boldsymbol{K}\boldsymbol{X} = \boldsymbol{0}. \quad (17)$$

The numerical results from this state-space approach are presented in the next Section in comparison with analytical and experimental results.

# 5 Results and discussion

## 5.1 Frequency band structures of a single column woodpile PC

Figure 8(a) shows the experimental measurements of the transmission gains obtained from a column of woodpile PC composed of 70 mm rods under 10.1 N pre-compression. We observe three pass bands in the test data separated by two stop bands. The magnitude of the first pass band is relatively small compared to the second and third pass bands. This is probably because the applied force in the low frequency level is relatively low in the experiment due to the interferences with the system setup.

The pass- and stop-bands obtained from the DEM simulation are presented in Fig. 8(b), showing good agreement with the experiment data. In this figure, we observe 21 peaks in each pass band, corresponding to the resonant frequencies of the finite chain. We also witness a pair of spikes in band gaps (see the inset of Fig. 8(b)). These peaks represent surface modes – a type of localized breathers – that are associated with the boundary conditions of lattice structures (Wallis, 1957). Specifically, the two-peak structure of the surface modes appears due to the splitting of the in-phase and out-of-phase vibrations of end particles that are constrained by the fixed boundaries (to be further explained in Section 5.2). Also, the two frequency of the two peaks are very close to each other, which indicate that they result from the near degeneracy of the two surface modes. Previous studies showed that these surface modes can be generated in diatomic chains exposed to free boundary conditions (Wallis, 1957; Theocharis et al., 2010). Other studies reported similar structures of localized breathers in monoatomic chains with light impurities (Theocharis et al., 2009; Starosvetsky et al., 2011). In this study, we find that the surface modes can be also generated in homogenous PCs, by leveraging their local resonances. Experimental results corroborate numerical findings, as we observe the spikes in the band gaps near the upper boundaries of the two high pass bands (marked in circles in Fig. 8(a)).

Figure 8(c) presents the dispersion curves obtained analytically from Eq. (9) under the assumption of an infinite chain. The vertical lines shown in Figs. 8(a), (b), and (c) denote the cutoff frequencies derived from the analytical dispersion relation. We find that the experimental, numerical, and analytical band structures are in excellent agreement. The inset images show the vibration modes of the unit cell, which trigger the generation of two upper pass-bands as discussed in Section 4.

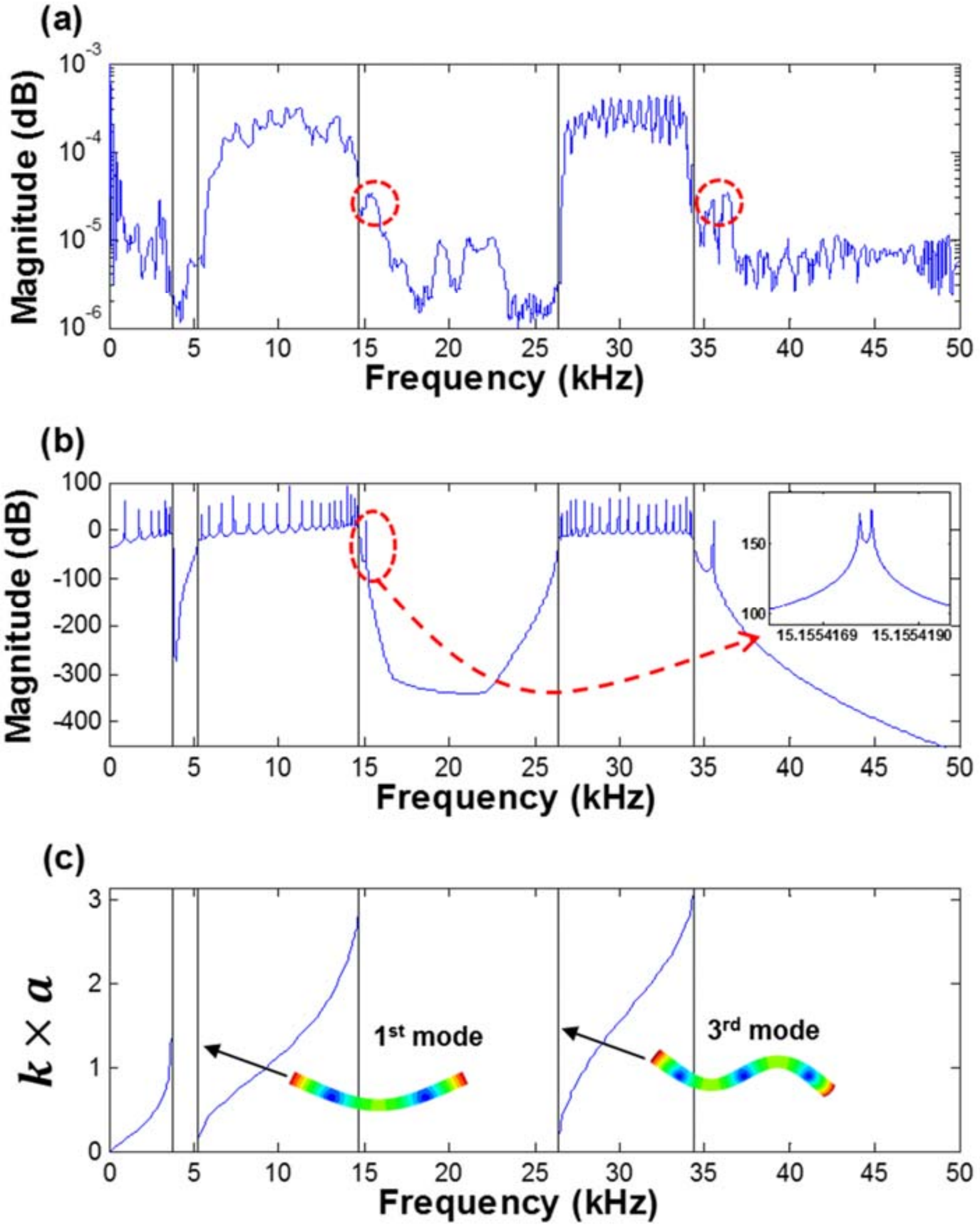

**Figure 8** Comparison of frequency responses of a single column woodpile PC consist of 70 mm-long rods (a) Experimental frequency spectrum. (b) Numerical frequency spectrum obtained from the DEM. Inset figure shows the two-peak structures in the second band gap due to surface modes. (c) Analytical dispersion curves of an infinite chain. The onset frequencies of the two upper pass bands correspond to the resonant frequencies of the 1st and 3rd bending modes of the cylinder.

## 5.2 Modal analysis

While the eigen-values represent the natural frequencies of the chain, the eigen-vectors denote its mode shapes at these resonant frequencies. The first mode shapes in three pass bands are presented in Figs 9 (a), (b), and (c). The vibration mode in the first band shows an oscillatory motion of cylinders with all three particles in each unit cell moving in phase. In the second pass band, we observe similar trends of long wavelength oscillations. However, we find that the auxiliary mass $m_1$ moves out of phase relative to the primary mass $M$ and the second auxiliary mass $m_2$ (Fig. 9(b)). This corresponds to the first bending mode of slender cylinders. In the third pass band, the primary mass $M$ and the secondary auxiliary mass $m_2$ move out of phase, while the first auxiliary mass $m_1$ remains stationary (Fig. 9(c)). This mode shape represents the second symmetric bending mode of the rods. Note that all three eigen-modes in Figs. 9(a)~(c) illustrate long wavelength oscillations with their wave number $k = \pi/aN$, where $a$ is equilibrium unit cell length, and $N$ is the number of rods in the chain.

As we increase frequencies in each pass band, we start to observe more complicated interactions between unit cells. For example, Fig. 9(d) shows the last mode shape in the second pass band. We find that the particles exhibit an out-of-phase mode shape with a shorter wave length. However, the relative motion of each mass in a unit cell remains similar to what we observed from the first resonant mode of the second band (compare Figs. 9(b) and 9(d)). Overall, as we move on to the higher mode of oscillations in a single pass band, we observe the oscillations with the shorter wavelength. Despite the change of wavelength, the particles in a unit cell keep the similar trends of oscillations.

In Figs. 9(e) and (f), we also present in-phase and out-of-phase surface modes positioned in the second band gap. We observe that these surface modes are characterized by the oscillatory particles in the boundaries that have relatively large displacements. The displacements of inner particles decrease roughly exponentially in the shape of $u_n = U(-1)^n e^{-\alpha n}$, where $n$ is the particle number from the boundary, $U$ is the displacement magnitude, and $\alpha$ is the attenuation coefficient (Wallis, 1957; Theocharis, 2010). Similar to the other resonant modes in the second pass band, the primary and the second auxiliary masses ($M$ and $m_2$) exhibit dominant motions in phase, while the first auxiliary mass ($m_1$) remains relatively stationary.

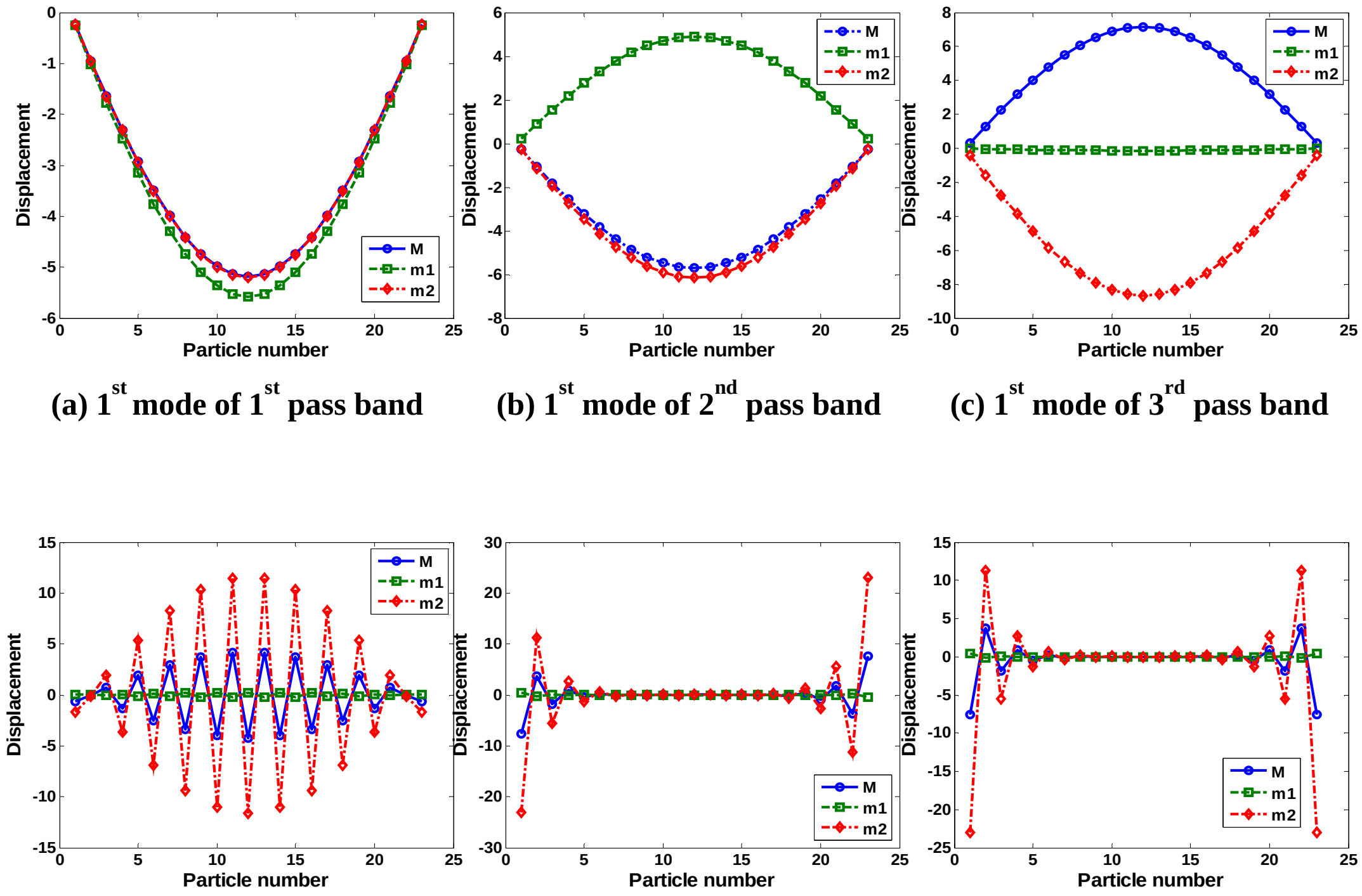

(a) 1st mode of 1st pass band (b) 1st mode of 2nd pass band (c) 1st mode of 3rd pass band

(d) 21st mode of 2nd pass band (e) Anti-symmetric surface mode (f) Symmetric surface mode

**Figure 9** The lowest-frequency mode shapes in the (a) first, (b) second, and (c) third pass bands. (d) The highest-frequency mode shape in the second pass band. (e) Out-of-phase surface mode in the second bandgap. (f) In-phase surface mode in the second bandgap.

## 5.3 Effect of pre-compression

We investigate the effect of pre-compression on the frequency response of the single column woodpile PC ($L$ = 70 mm). By using the DEM, we numerically obtain frequency spectra under different pre-compressions from 10.1 N to 39.9 N (Fig. 10). The empirical cutoff frequencies are also plotted in hollow square marks. In experiments, we determine the cutoff frequencies based on the criterion that the transmission gains at cutoff frequencies are 30% of the average gains in the corresponding pass-bands. We find that experimental results corroborate the numerical results well. A minute discrepancy is observed around the onset frequencies of the second and third pass bands ($f_{i2}$, $f_{i3}$), where experimental values are higher than the corresponding simulation results. Up-shift of experimental results has been observed in previous studies, and several possible reasons of such discrepancies are summarized in the reference (Boechler et al., 2010).

In Fig. 10, we note that the upper cutoff frequencies of each pass band ($f_{c1}, f_{c2}, f_{c3}$) increase as we impose the larger pre-compression. However, we observe that the lower cutoff frequencies ($f_{i1}, f_{i2}, f_{i3}$) are not affected by the pre-compression. As stated earlier, this is because the lower cutoff frequencies of each band are resonant frequencies of a standalone rod. The experimental results confirm the validity of our numerical simulations based on the DEM. In the map, the formation of surface modes can also be observed in the forbidden frequency band above the topmost pass band (indicated by an arrow). The surface modes in the lower band gap are not clearly seen due to their proximity to the cutoff frequencies.

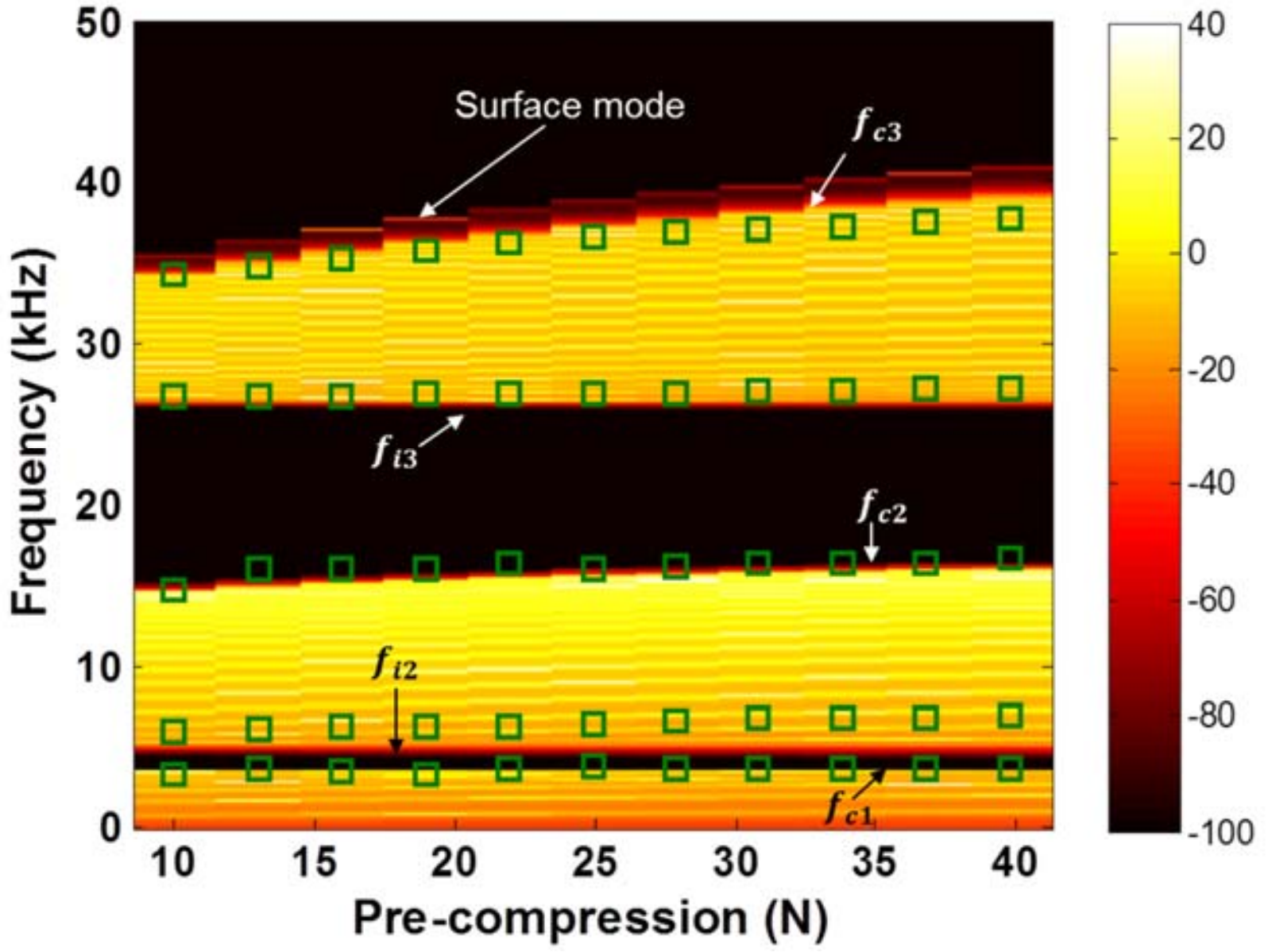

**Figure 10** Surface map of wave transmission in a woodpile PC as a function of frequency and pre-compression. The color intensity denotes numerically calculated transmission gains in decibel. The cutoff frequencies measured from experiments are marked in hollow squares.

### 5.4 Effect of cylinder's length

Lastly we examine how a cylinder's length affects the frequency band structure in single column woodpile PCs. For numerical considerations, we first run the optimization scheme to find mass and spring parameters in DEM as described in Sec. 4.1. As a result, the optimized mass distributions are summarized in Table 2. We find that as the length of the rod increases, the mass ratios of the primary mass ($M$) and the second auxiliary mass ($m_2$) increase, while the mass ratio of the first auxiliary mass ($m_1$) decreases. Based on the optimal mass distributions, the frequency responses of finite woodpile chains with various rod lengths are simulated via the state-space approach. Figure 11 illustrates the numerical results of transmission gains along with the cutoff frequencies measured in experiments (square marks). We find that the numerical and experimental results show excellent agreement. As the lengths of rods increase, we observe pass- and stop-bands shift to lower frequencies accordingly. This trend is predictable by that the lower cutoff frequencies of pass bands are governed by the bending resonant frequencies of rods, which decrease as the rod length increases. According to the Euler beam theory, these bending frequencies are inversely proportional to the square of rod length (i.e., $f \propto 1/L^2$ ). The fitted curves of this inverse relationship are shown in Fig. 11. We find that the onset frequencies of 2nd and 3rd pass bands (i.e., $f_{i2}$ and $f_{i3}$) are well matched with these fitted curves (mathematically, ${C_I}/{L^2}$ and ${C_{II}}/{L^2}$, where fitting parameters are $C_I = 2{,}7480$ and $C_{II} = 128{,}740$).

**Table 2** Optimized mass distributions for various woodpile PCs composed of identical cylinders with their lengths $L$ = 40, 50, 60, 70, and 80 mm.

| Mass (g) | 40mm | 50mm | 60mm | 70mm | 80mm |
|---|---|---|---|---|---|
| Total mass | 1.732 | 2.165 | 2.598 | 3.031 | 3.464 |
| M<br>(mass ratio) | 0.883<br>(0.51) | 0.422<br>(0.19) | 0.647<br>(0.25) | 0.850<br>(0.28) | 1.171<br>(0.34) |
| $m_1$<br>(mass ratio) | 0.849<br>(0.49) | 1.420<br>(0.66) | 1.469<br>(0.56) | 1.413<br>(0.47) | 1.232<br>(0.36) |
| $m_2$<br>(mass ratio) | | 0.323<br>(0.15) | 0.482<br>(0.19) | 0.768<br>(0.25) | 1.061<br>(0.30) |

The effect of rod length variations on frequency band shifts is more drastic compared to the tunability obtained from the pre-compression changes (compare Figs. 10 and 11). For example, the upper cutoff frequencies of the second highest pass bands are reduced approximately by a half, when the rod length changes from 40 mm to 80 mm. This implies that we can tailor the responses of woodpile

PCs over wide frequency spectra by manipulating spatial dimensions of the woodpile PC architectures. Such drastic effects have been also observed when we change the contact angles of cylindrical elements in a chain of PCs, though we do not include the results for the brevity of the manuscript. By combining the effects of these different design parameters, we expect to achieve a versatile frequency filtering system based on woodpile PCs.

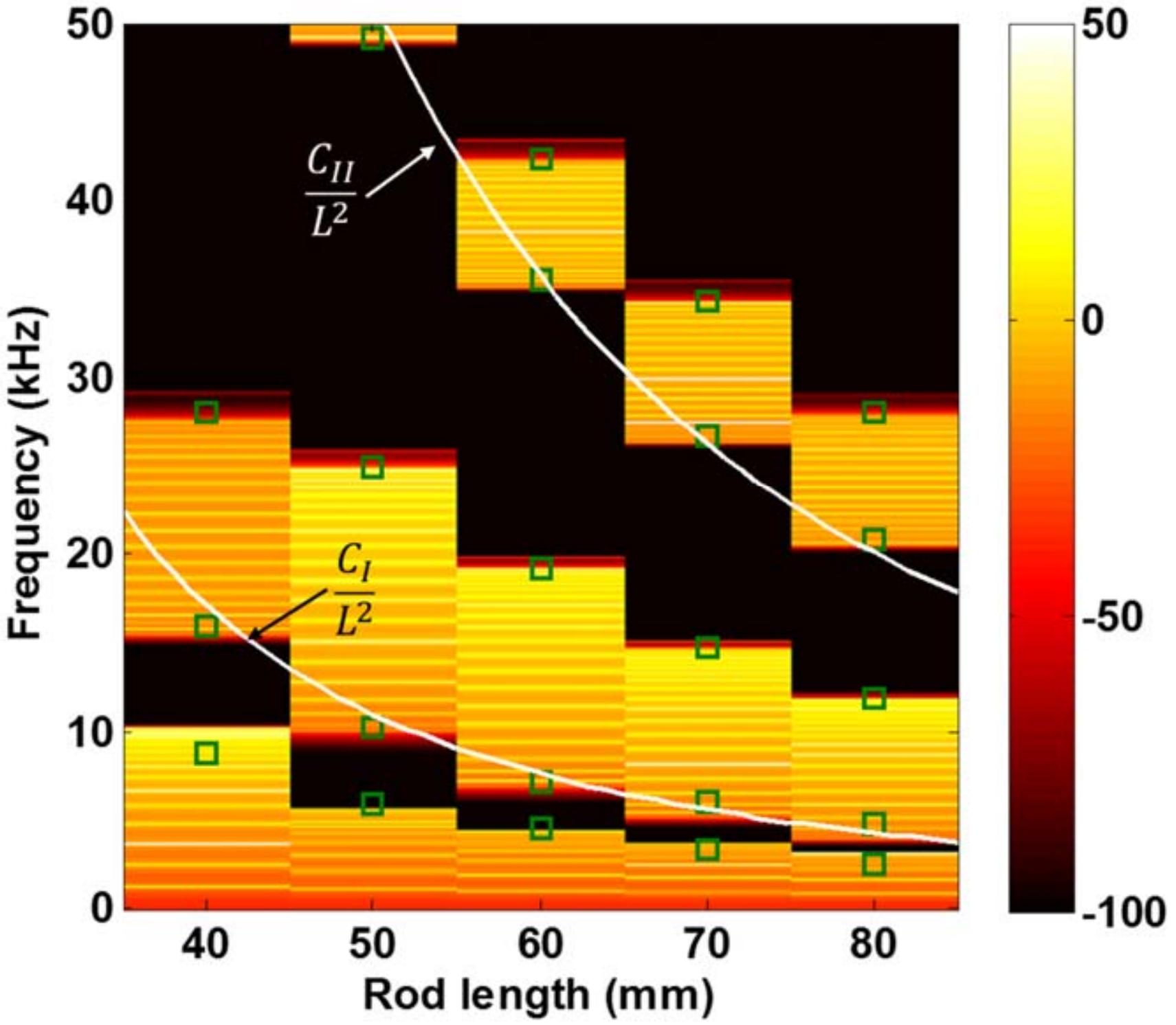

**Figure 11** Frequency spectra of woodpile PCs as a function of the rod length under 10.1 N pre-compression. The fitted white curves are based on the relationship between the rods' bending frequencies and the rod length ($\omega \propto 1/L^2$ ).

## 6 Conclusion

In this study, we investigated the formation of frequency band structures in single column woodpile PCs consist of slender cylindrical rods. We experimentally found that the bending modes of the constituent rods significantly affect the dispersive responses of elastic waves that propagate along the stacking direction of the woodpile PCs. This hints that we can achieve the construction of locally resonant phononic crystals by means of simple, homogeneous architectures of woodpile PCs without relying on heterogeneous constituents composed of heavy oscillators and soft enclosures. This offers a significant merit in designing vibration filtering devices, especially in a relatively low frequency domain.

To describe the wave dispersion mechanism, we built a discrete element model, whose mass and stiffness parameters were determined empirically via an optimization scheme. Using this model, we verified that the symmetric vibration modes of cylinders significantly affect the dispersion mechanisms of axial waves, while the anti-symmetric modes do not transmit along the stacking direction of the woodpile PCs. In addition, the lower cutoff frequencies of pass bands are governed by the natural frequencies of rods' bending vibration modes. These findings can provide us with useful insights for the design of vibration filtering devices in the form of woodpile PCs. This study has been limited to the linear responses of single column woodpile PCs. However, the numerical and experimental results of this study can be further extended to the studies of fully stretched woodpile PCs. The formation, modulation, and attenuation mechanisms of nonlinear waves in woodpile PCs will be reported in the authors' upcoming publication.

## Acknowledgement

The authors acknowledge support from the National Science Foundation (Grant No. 1234452). We also thank Drs. Chiara Daraio, Panayotis Kevrekidis, and Devvrath Khatri for their helpful input.

## References

Boechler, N., Theocharis, G., Job, S., Kevrekidis, P.G., Porter, M.A., Daraio, C., 2010. Discrete breathers in one-dimensional diatomic granular crystals. Phys. Rev. Lett. 104, 244302.

Boechler. N., Yang, J., Theocharis, G., Kevrekidis, P.G., Daraio, C., 2011. Tunable vibrational band gaps in one-dimensional diatomic granular crystals with three-particle unit cells. J. Appl. Phys. 109, 074906.

Brillouin, L., 1953. Wave propagation in periodic structures. New York: Dover

Franklin, G.F., Emami-Naeini, A., Powell, J.D., 1993. Feedback control of dynamic systems: Addison-Wesley Longman Publishing Co., Inc.

Herbold, E., Kim, J., Nesterenko, V., Wang, S., Daraio, C., 2009. Pulse propagation in a linear and nonlinear diatomic periodic chain: effects of acoustic frequency band-gap. Acta Mech. 205, 85-103.

Hirsekorn, M., 2004. Small-size sonic crystals with strong attenuation bands in the audible frequency range. Appl. Phys. Lett. 84, 3364-3366.

Hussein, M.I., Hulbert, G.M., Scott, R.A., 2007. Dispersive elastodynamics of 1D banded materials and structures: Design. J. Sound Vib. 307, 865-93.

Jiang, H., Wang, Y., Zhang, M., Hu, Y., Lan, D., Zhang, Y., Wei, B., 2009. Locally resonant phononic woodpile: A wide band anomalous underwater acoustic absorbing material. Appl. Phys. Lett. 95, 104101-3.

Johnson, K.L. 1985. Contact Mechanics. Cambridge University Press.

Khatri, D., Ngo, D., Daraio, C., 2012. Highly nonlinear solitary waves in chains of cylindrical particles. Granul. Matter 14, 63-69.

Kushwaha, M.S., 1996. Classical band structure of periodic elastic composites. Int. J. Mod. Phys. B (Singapore). 10, 977-1094.

Li, F., Ngo, D., Yang, J., Daraio, C., 2012. Tunable phononic crystals based on cylindrical Hertzian contact. Appl. Phys. Lett. 101, 171903.

Lin, S.Y., Fleming, J.G., Hetherington, D.L., Smith, B.K., Biswas, R., Ho, K.M., Sigalas, M.M., Zubrzycki, W., Kurtz, S.R., Bur, J., 1998. A three-dimensional photonic crystal operating at infrared wavelengths. Nature 394, 251-253.

Liu, Z.Y., Zhang, X.X., Mao, Y.W., Zhu, Y.Y., Yang, Z.Y., Chan, C.T., Sheng P., 2000. Locally resonant sonic materials. Sci. 289,1734-6.

Noda, S., Tomoda, K., Yamamoto, N., Chutinan, A., 2000. Full three-dimensional photonic bandgap crystals at near-infrared wavelengths. Science 289, 604-606.

Puttock, M.J., Thwaite, E.G., 1969. Elastic compression of spheres and cylinders at point and line contact. Natl. Stand. Lab. Tech. Pap. 25, Commonwealth Scientific and Industrial Research Organization, Australia.

Starosvetsky, Y., Jayaprakash, K. R., Vakakis, A. F., 2011, Scattering of solitary waves and excitation of transient breathers in granular media by light intruders and no precompression, Journal of Applied Mechanics, 79, 011001.

Theocharis, G., Kavousanakis, M., Kevrekidis, P. G., Daraio, C., Porter, M. A., Kevrekidis, I. G., 2009, Localized breathing modes in granular crystals with defects, Phys. Rev. E, 80, 066601.

Theocharis, G., Boechler, N., Kevrekidis, P.G., Job, S., Porter, M.A., Daraio, C., 2010. Intrinsic energy localization through discrete gap breathers in one-dimensional diatomic granular crystals. Phys. Rev. E, 82, 056604.

Vasseur, J.O., Deymier. P.A., Chenni, B., Djafari-Rouhani, B., Dobrzynski, L., Prevost, D., 2001. Experimental and theoretical evidence for the existence of absolute acoustic band gaps in two-dimensional solid phononic crystals. Phys. Rev. Lett. 86, 3012-3015.

Wallis, R.F., 1957. Effect of free ends on the vibration frequencies of one-dimensional lattices, Phys. Rev. 105(2), 540-545.

Wu, L.Y., Chen, L.W., 2011. Acoustic band gaps of the woodpile sonic crystal with the simple cubic lattice. J. Phys. D: Appl. Phys. 44, 045402.

Yang, J., Dunatunga, S., Daraio, C., 2012. Amplitude-dependent attenuation of compressive waves in curved granular crystals constrained by elastic guides. Acta Mech. 223, 549-562.